\documentclass[submission,copyright,creativecommons]{eptcs}
\providecommand{\event}{F-IDE 2016} 
\usepackage{underscore}           

\usepackage{hyperref}

\usepackage[setrelation]{math}
\usepackage{amsmath}
\usepackage{amsthm}
\usepackage[modernsign,substopindex,shortmquant,mquantifiertype,
mconnectiveformal,postfixflatinterpret,bracketmodalinterpret,
setfixinterpret,modifopindex,seqinfers,seqoptional,footnotecalculus,abbrseqcontext,shortterms,nosigmaterms,novarterms]{logic}
\usepackage[pretest,nocommandblocks]{progreg}
\usepackage[bracketinterpret,postfixinterpret,bracketmodalinterpret,simplenames]{dL}
\usepackage{tabularx}
\usepackage{booktabs}
\usepackage{paralist}
\usepackage{nfplt}
\usepackage{todonotes}
\usepackage{wrapfig}
\usepackage{float}
\usepackage{graphics}
\usepackage{caption}
\usepackage{subfig}
\usepackage{enumitem}

\usepackage{xcolor}
\usepackage{color}
\definecolor{gray}{rgb}{0.5,0.5,0.5}

\usepackage{prettyref}
\newcommand{\rref}[2][]{\prettyref{#2}}
\newrefformat{model}{Model\,\ref{#1}}
\newrefformat{listing}{Listing\,\ref{#1}}
\newrefformat{line}{line\,\ref{#1}}
\newrefformat{sec}{Section\,\ref{#1}}
\newrefformat{appendix}{Appendix\,\ref{#1}}
\newrefformat{def}{Definition\,\ref{#1}}
\newrefformat{defn}{Definition\,\ref{#1}}
\newrefformat{thm}{Theorem\,\ref{#1}}
\newrefformat{ax}{\ref{#1}}
\newrefformat{prop}{Proposition\,\ref{#1}}
\newrefformat{lem}{Lemma\,\ref{#1}}
\newrefformat{cor}{Corollary\,\ref{#1}}
\newrefformat{ex}{Example\,\ref{#1}}
\newrefformat{tab}{Table\,\ref{#1}}
\newrefformat{fig}{Figure\,\ref{#1}}
\newrefformat{eqn}{(\ref{#1})}
\newrefformat{proof}{Proof\,\ref{#1}}

\usepackage{listings}
\usepackage{verbatim}
\usepackage{courier} 
\usepackage{upquote} 

\lstdefinelanguage{scala}{
  morekeywords={abstract,case,catch,class,def,%
    else,extends,false,final,finally,%
    for,if,implicit,import,match,mixin,%
    new,null,object,override,package,%
    private,protected,requires,return,sealed,%
    super,this,throw,trait,true,try,%
    type,val,var,while,with,yield},
  otherkeywords={=>,<-,<\%,<:,>:,\#,@,by,&,|,<,?},
  sensitive=true,
  morecomment=[l]{//},
  morecomment=[n]{/*}{*/},
  morestring=[b]",
  morestring=[b]"""
}

\lstdefinelanguage{bellerophon}{
  morekeywords={partial,OnAll,'R,'L,'_},
  otherkeywords={&,|,<,?},
  sensitive=true,
  morecomment=[l]{//},
  morecomment=[n]{/*}{*/},
  morestring=[b]",
  morestring=[b]"""
}

\floatstyle{ruled}
\newfloat{model}{tbp}{model}
\floatname{model}{Model}

\newcommand{\memnode}[2][right]{\tikz[remember picture,overlay]\node[#1] (#2) {};}
\newcommand{\connectnodes}[2]{\tikz[remember picture,overlay]
\draw[->] (#1) -- (#2);}

\graphicspath{{./figs/}}

\title{The \KeYmaeraX Proof IDE\\\normalsize{Concepts on Usability in Hybrid Systems Theorem Proving}}
\author{Stefan Mitsch 
\institute{
  Computer Science Department, Carnegie Mellon University, Pittsburgh, PA, USA
}
\email{smitsch@cs.cmu.edu}
\and 
Andr\'{e} Platzer
\institute{
  Computer Science Department, Carnegie Mellon University, Pittsburgh, PA, USA
}
\email{aplatzer@cs.cmu.edu}
}
\def\titlerunning{The \KeYmaeraX Proof IDE}
\def\authorrunning{S. Mitsch \& A. Platzer}

\begin{document}

\maketitle

\begin{abstract}
Hybrid systems verification is quite important for developing correct controllers for physical systems, but is also challenging.
Verification engineers, thus, need to be empowered with ways of guiding hybrid systems verification while receiving as much help from automation as possible.
Due to undecidability, verification tools need sufficient means for intervening during the verification and need to allow verification engineers to provide system design insights.

This paper presents the design ideas behind the user interface for the hybrid systems theorem prover \KeYmaeraX.
We discuss how they make it easier to prove hybrid systems as well as help learn how to conduct proofs in the first place.
Unsurprisingly, the most difficult user interface challenges come from the desire to integrate automation and human guidance.
We also share thoughts how the success of such a user interface design could be evaluated and anecdotal observations about it.
\end{abstract}

\section{Introduction}

\emph{Cyber-physical systems} such as cars, aircraft, and robots combine computation and physics, and provide exceedingly interesting and important verification challenges.
\KeYmaeraX \cite{DBLP:conf/cade/FultonMQVP15} is a theorem prover for \emph{hybrid systems}, i.\,e., systems with interacting discrete and continuous dynamics, which arise in virtually all mathematical models of cyber-physical systems.\footnote{\KeYmaeraX is available at \url{http://keymaeraX.org/}}
It implements \emph{differential dynamic logic} (\dL \cite{DBLP:journals/jar/Platzer08,DBLP:conf/lics/Platzer12a,DBLP:journals/jar/Platzer16}) for \emph{hybrid programs}, a program notation for hybrid systems.
Differential dynamic logic provides compositional techniques for proving properties about hybrid systems.
Despite the substantial advances in automation, user input is often quite important, since hybrid systems verification is not semidecidable \cite{DBLP:journals/jar/Platzer08}.
Human insight is needed most notably for finding invariants for loop induction and finding differential invariants for unsolvable differential equations \cite{DBLP:conf/lics/Platzer12a}.
But even some of the perfectly decidable questions in hybrid systems verification are intractable in practice,
such as the final step of checking the validity of formulas in real arithmetic\footnote{Decision procedures are doubly exponential in the number of quantifier alternations \cite{DBLP:journals/jsc/DavenportH88}, and practical implementations  doubly exponential in the number of variables.} in a \dL proof.

To overcome those verification challenges, \KeYmaeraX combines automation and interaction capabilities to enable users to verify their applications even if they are still out of reach for state-of-the-art automation techniques.
The central question in usability, thus, is how interaction and automation can jointly solve verification challenges.
For isolated strategic aspects of the proofs, such user guidance is easily separated, for example when using system insights to provide loop invariants and differential invariants.
Other aspects of human insights are more invasive, such as picking and transforming relevant formulas to make intractable arithmetic tractable.
Users have to link the logical level of proving (e.\,g., the conjecture, available proof rules and axioms) to the abstract interaction level (e.\,g., visualization of formulas in a sequent) and decide about concrete interaction steps (e.\,g., where to click, what to type) \cite{DBLP:journals/jsc/AitkenGMT98}.
Hence, going back and forth between automated proof tactics and user guidance poses several challenges:
\begin{itemize}[noitemsep,nolistsep]
\item Users need to be provided with a way of understanding the proof state produced by automated tactics. What are the open proof goals? How are these goals related to the proof of the conclusion? And why did the automated tactic stop making progress?
\item Users need to be provided with efficient ways of understanding the options for making progress. What tactics are available and where can they be applied? What input is needed? And what interactions provide genuinely new insights into a proof that the automation would not have tried?
\item Users need efficient tools for executing proof steps. How to provide gradual progress for novices? How to let experienced users operate with minimal input? How to reuse proof steps across similar goals? And how to generalize a specific tactic script into a proof search procedure for similar problems?
\end{itemize}
Users may additionally benefit from picking an appropriate interaction paradigms, such as \emph{proof as programming}, \emph{proof by pointing}, or \emph{proof by structure editing} \cite{DBLP:journals/jsc/AitkenGMT98}.

This paper discusses how \KeYmaeraX addresses these challenges through its web-based user interface.
The complexity of larger hybrid systems verification challenges require that users be granted significant control over how a proof is conducted and what heuristics are applied for proof search.
To this end, \KeYmaeraX separates user interaction and proof search from the actual proof steps in the prover kernel to ensure soundness while providing reasoning flexibility \cite{DBLP:conf/cade/FultonMQVP15}. 
All proof steps follow from a small set of axioms by uniform substitution \cite{DBLP:conf/cade/Platzer15,DBLP:journals/jar/Platzer16}.
%
This paper further introduces an evaluation concept to determine how effectively alternative user interaction concepts implemented in \KeYmaeraX address these challenges, as well as whether or not the combination of these concepts compare favorably to our previous hybrid systems theorem prover \KeYmaera \cite{DBLP:conf/cade/PlatzerQ08}.
The experiments are yet to be conducted; our reports on the effectiveness of the user interaction concept remain anecdotal based on feedback from external users and students.

\section{Preliminaries: Differential Dynamic Logic}
\label{sec:preliminaries}

This section recalls the syntax and semantics of differential dynamic logic by example of the motion of a person on an escalator.
This example serves for illustrating prover interaction throughout the paper.

\paragraph{Syntax and semantics by example.}

Suppose a person is standing on an escalator, which moves upwards with non-negative speed \(v{\geq}0\), so the person's vertical position \(x\) follows the differential equation \(\D{x}=v\).
If the person is not at the bottom-most step (\(\ptest{x{>}1}\)), she may step down one step (\(\humod{x}{x-1}\)) or may just continue moving upwards; since she may or may not step down, even if allowed, we use a non-deterministic choice (\(\cup\)).
We want to prove that the person never falls off the bottom end of the escalator (\(x{\geq}0\)) when stepping down and moving upwards are repeated arbitrarily often (modeled with the repetition operator \(\prepeat{}\)).

\begin{equation}\label{eq:example}
\underbrace{x \geq 2 \land v \geq 0 \limply}_\text{initial conditions} \dbox{\underbrace{\prepeat{\bigl((\pchoice{\ptest{x>1};\humod{x}{x-1})}{\D{x}=v}\bigr)}}_\text{hybrid program}}\underbrace{x \geq 0}_{\mathclap{\text{safety condition}}}
\end{equation}

The formula \eqref{eq:example} captures this example as a safety property in \dL. Suppose, the person is initially at some position $x\geq2$ when the escalator turns on. From any state satisfying the initial conditions \(x \geq 2 \land v \geq 0\), all runs of the hybrid program \(\prepeat{(\pchoice{(\ptest{x>1};\humod{x}{x-1})}{\D{x}=v})}\) reach only states that satisfy the safety condition \(x \geq 0\).
More detailed examples on modeling hybrid systems are in \cite{DBLP:journals/sttt/QueselMLAP16}.

\paragraph{Manual proof in \dL.}

Formulas in \dL, such as the simple example \eqref{eq:example}, can be proved with the \dL proof calculus.
Proofs in \dL are sequent proofs: a sequent has the shape \(\lsequent{\Gamma}{\Delta}\), where we assume all formulas \(\Gamma\) in the antecedent (to the left of the turnstile \(\lsequent{{}}{{}}\)) to show any of the formulas \(\Delta\) in the succedent (right of the turnstile).
The sequent notation works from the desired conclusion at the bottom toward the resulting subgoals at the top.
While sometimes surprising for novices, this notation emphasizes how the truth of the conclusions follows from the truth of their respective premises top-down.
This notation also highlights the current subquestion at the very top of the deduction.
Steps in the sequent proof are visualized through a horizontal line, which separates the conclusion at the bottom from the premises at the top.
The name of the deduction step is annotated to the left of the horizontal line.
For example, the proof rule \(\dbox{\cup}\) says that to conclude safety of a non-deterministic choice of programs \(\pchoice{\alpha}{\beta}\) (below the bar) it suffices to prove safety of both \(\alpha\) and \(\beta\) individually (above bar).

\irlabel{choiceb|$[\cup]$}
\begin{sequentdeduction}[array]
\linfer[choiceb]
{
\lsequent{\phi}{\dbox{\alpha}\psi \land \dbox{\beta}\psi}
}
{\lsequent{\phi}{\dbox{\pchoice{\alpha}{\beta}}\psi}}
\end{sequentdeduction}

\begin{proof}[Manual proof example]
\label{proof:example}
Starting from the formula \eqref{eq:example} at the very bottom of the deduction, we develop a safety proof as follows.
As first step, the rule \(\limply\textsf{R}\) moves the assumptions from the left-hand side of an implication into the antecedent. 
Next \(\land\textsf{L}\) splits the conjunction \(x \geq 2 \land v \geq 0\) into individual facts (i.\,e., we get to assume both, \(x \geq 2\) and \(v \geq 0\) individually).
Then, we use loop induction with the invariant \(x > 0\).
We have to show three cases: the loop invariant must hold initially (base case \(x \geq 2 \limply x>0\)), it must be strong enough to entail safety (use case \(x>0 \limply x \geq 0\)), and it must be preserved by the loop body (induction step \(\dbox{\pchoice{(\ptest{x>1};\humod{x}{x-1})}{\D{x}=v}}x > 0\)).

\irlabel{composeb|$[{;}]$}
\irlabel{testb|$[?]$}
\irlabel{assignb|$[:=]$}
\irlabel{implyR|$\limply$R}
\irlabel{andR|$\land$R}
\irlabel{andL|$\land$L}
\irlabel{loop|loop}
\irlabel{ode|ODE}
\irlabel{QE|QE}

\vbox{\renewcommand{\linferPremissSeparation}{~~~}%
\begin{sequentdeduction}[array]
\linfer[andR]
{
\linfer[QE]
{\lclose}
{
\linfer[assignb]
{\lsequent{x{>}0, v{\geq}0, x{>}1}{x-1{>}0}}
{
\linfer[testb+implyR]
{\lsequent{x{>}0, v{\geq}0, x{>}1}{\dbox{\humod{x}{x-1}}x{>}0}}
{
\linfer[composeb]
{\lsequent{x{>}0, v{\geq}0}{\dbox{\ptest{x{>}1}}\dbox{\humod{x}{x-1}}x {>}0}}
{
\lsequent{x{>}0, v{\geq}0}{\dbox{\ptest{x{>}1};\humod{x}{x-1}}x{>}0}
}
}
}
}
!
\linfer[QE]
{\lclose}
{
\linfer[ode]
{\lsequent{x_0{>}0, v{\geq}0}{\forall t{\geq}0\, \forall 0{\leq}s{\leq}t~( x=x_0+vs \limply x{>}0)}}
{
\lsequent{x{>}0, v{\geq}0}{\dbox{\D{x}=v}x{>}0}
}
}
}
{
\linfer[choiceb]
{\lsequent{x{>}0, v{\geq}0}{\dbox{\ptest{x{>}1};\humod{x}{x-1}}x{>}0 \land \dbox{\D{x}=v}x{>}0}}
{
\ldots \memnode{inductionstepproof}
}
}
\end{sequentdeduction}

\begin{sequentdeduction}[array]
\linfer[loop]
{
\linfer[QE]
{\text{(base case)}~\lclose}
{
\lsequent{x{\geq}2, v{\geq}0}{x{>}0}
}
!
\linfer[QE]
{\text{(use case)}~\lclose}
{
\lsequent{x{>}0}{x{\geq}0}
}
!
\linfer
{\memnode{inductionstepstart}~\text{(induction step)}}
\lsequent{x{>}0, v{\geq}0}{\dbox{\pchoice{(\ptest{x{>}1};\humod{x}{x-1})}{\D{x}=v})}x{>}0}
}
{
\linfer[andL]
{\lsequent{x{\geq}2, v{\geq}0}{\dbox{\prepeat{(\pchoice{(\ptest{x{>}1};\humod{x}{x-1})}{\D{x}=v})}}x{\geq}0}}
{
\linfer[implyR]
{\lsequent{x{\geq}2 \land v{\geq}0}{\dbox{\prepeat{(\pchoice{(\ptest{x{>}1}; \humod{x}{x-1})}{\D{x}=v})}}x{\geq}0}}
{\lsequent{{}}{x{\geq}2 \land v{\geq}0 \limply \dbox{\prepeat{(\pchoice{(\ptest{x{>}1};\humod{x}{x-1})}{\D{x}=v})}}x{\geq}0}
}
}
}
\end{sequentdeduction}

\connectnodes{inductionstepstart}{inductionstepproof}
}

Here, the base case and use case can be shown easily using quantifier elimination \textsf{QE}.
The induction step proceeds using the proof rule \(\dbox{\cup}\) for non-deterministic choice that we saw above, followed by \(\land\textsf{R}\) to split the induction step proof into two branches. 
On the first branch, we first turn the sequential composition (\(;\)) into nested boxes (\(\dbox{\ptest{x>1}}\dbox{\humod{x}{x-1}}x>0\)), and then use the test condition (\(x>1\)) as an additional assumption using rule \(\dbox{\ptest{}}\) followed by \(\limply\textsf{R}\). 
We show safety of the assignment \(\humod{x}{x-1}\) using quantifier elimination \textsf{QE} to close the proof.
On the second branch, we show safety of the differential equation \(\D{x}=v\) using the proof rule \textsf{ODE} followed by \textsf{QE}.
\end{proof}

\section{\KeYmaeraX Proof Automation}

As a basis for understanding how \KeYmaeraX searches for proofs and where and why it asks for user guidance, this section gives a high-level explanation of \KeYmaeraX tactics. 

\KeYmaeraX automates the tedious task of proving steps that follow unambiguously from the structure of the conjecture.
It further provides (heuristic) tactics to generate and explore invariant candidates for loop induction and differential equations.
One might imagine \KeYmaeraX to try to solve differential equations and use the solution to guide a differential invariant proof, before it resorts to more involved differential invariant proofs.
\KeYmaeraX provides proof tactics for propositional reasoning, reasoning about hybrid programs, and closing (arithmetic) proof goals, which are combined into a fully automated proof search tactic.
For example, with a loop invariant candidate annotated in the \KeYmaeraX input file, the running example in this paper proves fully automated.


Propositional reasoning \textsf{prop} and program unfolding \textsf{unfold} of hybrid programs follows along propositional sequent rules and the axioms of \dL. 
These tactics successively match on the shape of a formula to transform it into simpler parts, before the tactics descend into the resulting parts.
Program unfolding focuses on the decidable fragment of reasoning about hybrid programs: it stops and asks for user guidance when it encounters loops or ODEs.
The following proof snippet applies \textsf{unfold} to just the induction step of the escalator proof.

\begin{proof}[Induction step by \textsf{unfold}]
\label{proof:unfold}
The tactic \textsf{unfold} applies hybrid program axioms and splits conjunctions in the succedent into proof branches, but stops when it encounters loops or ODEs.
The resulting two subgoals correspond to the two (logically unfolded) paths through our running example: we have to show safety of the discrete assignment \(\humod{x}{x-1}\) as well as of the differential equation \(\D{x}=v\).
\irlabel{unfold|unfold}
\begin{sequentdeduction}[array]
\linfer[unfold]
{
\lsequent{x{>}0, v{\geq}0, x{>}1}{x-1 \geq 0}
!
\lsequent{x{>}0, v{\geq}0}{\dbox{\D{x}=v}x{>}0}
}
{\lsequent{x{>}0, v{\geq}0}{\dbox{\pchoice{\ptest{x{>}1};\humod{x}{x-1}}{\D{x}=v}}x{>}0}}
\end{sequentdeduction}
\end{proof}




\KeYmaeraX ships with proof search tactic \textsf{auto}, which combines propositional reasoning with program unfolding, loop invariant exploration, certain automated proof techniques for differential equations, and proof closing by quantifier elimination.
Even though the \textsf{auto} tactic finds proofs for important classes of hybrid systems automatically, it still may stop exploration and ask for user guidance in complicated cases (e.\,g., when none of the explored differential invariants helps closing the proof).


\section{\KeYmaeraX User Interaction}
\label{sec:ui}

When the automated tactics shipped with \KeYmaeraX fail to find a proof (due to a wrong model, missing loop or differential invariants, or intractable arithmetic), user interaction is needed to improve the model and make progress with the proof.
This section introduces the \KeYmaeraX user interaction concepts and their implementation in a graphical web-based user interface.
The user interface of \KeYmaeraX is based on these principles and hypotheses:
\begin{description}
\item[Familiarity] Prover user interfaces benefit from a familiar look\&feel that resembles how proofs are conducted in theoretical developments and that are compatible with the way that proof rules are presented. 
\item[Traceability] Sophisticated verification challenges, especially in hybrid systems, need a way of mixing automation with user guidance in a way that the user can trace and understand the respective remaining questions.
\item[Tutoring] Interactive proof-by-pointing at formulas and terms are an efficient way of learning how to conduct proofs and help internalizing reasoning principles by observation.
Tactic recording is an efficient way of learning how to write tactics by observing to which interactive proof steps they correspond.
\item[Flexibility] Humans reason in more flexible ways than automation procedures. User interfaces should allow proof steps at any part of a proof as well as free-style transformations of formulas, and they should embrace multiple reasoning styles, such as explicit proofs, proof-by-pointing, and proof-by-search.
\item[Experimentation] A strict separation of prover core and prover interface not only helps soundness, but also enables more agile experimentation with new styles of conducting proofs and of interacting with provers.
\end{description}

\begin{figure}[t]
\begin{tikzpicture}
\node at (0,0) {\includegraphics[width=.98\columnwidth]{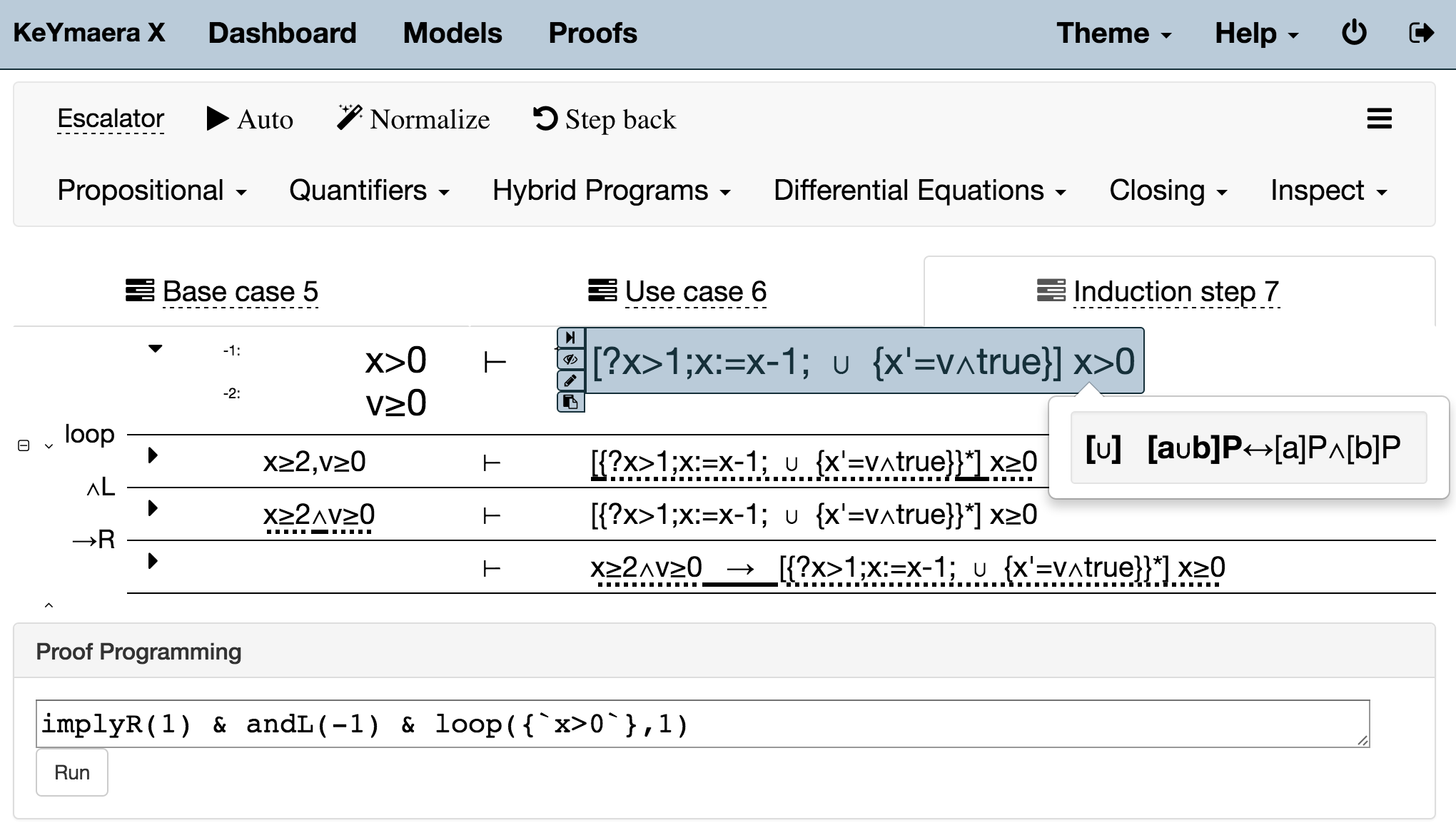}};
\node[draw,fill=black!10,circle] at (-7.5,2.5) {\bf{1}};
\node[draw,fill=black!10,circle] at (-7.5,0.2) {\bf{2}};
\node[draw,fill=black!10,circle] at (-4.5,-2.6) {\bf{3}};
\node[draw,fill=black!10,circle] at (6.7,1.4) {\bf{4}};
\node[draw,fill=black!10,circle] at (7.5,0.2) {\bf{5}};
\node[draw,fill=black!10,circle] at (6,-1.7) {\bf{6}};
\end{tikzpicture}
\caption{Screenshot of \KeYmaeraX with annotated user interface elements: \textcircled{1} proof-by-search; \textcircled{2} sequent view; \textcircled{3} proof programming and tactic extraction; \textcircled{4} proof branch; \textcircled{5} tactic suggestion; \textcircled{6} tactic step highlighting.}
\label{fig:overview}
\end{figure}
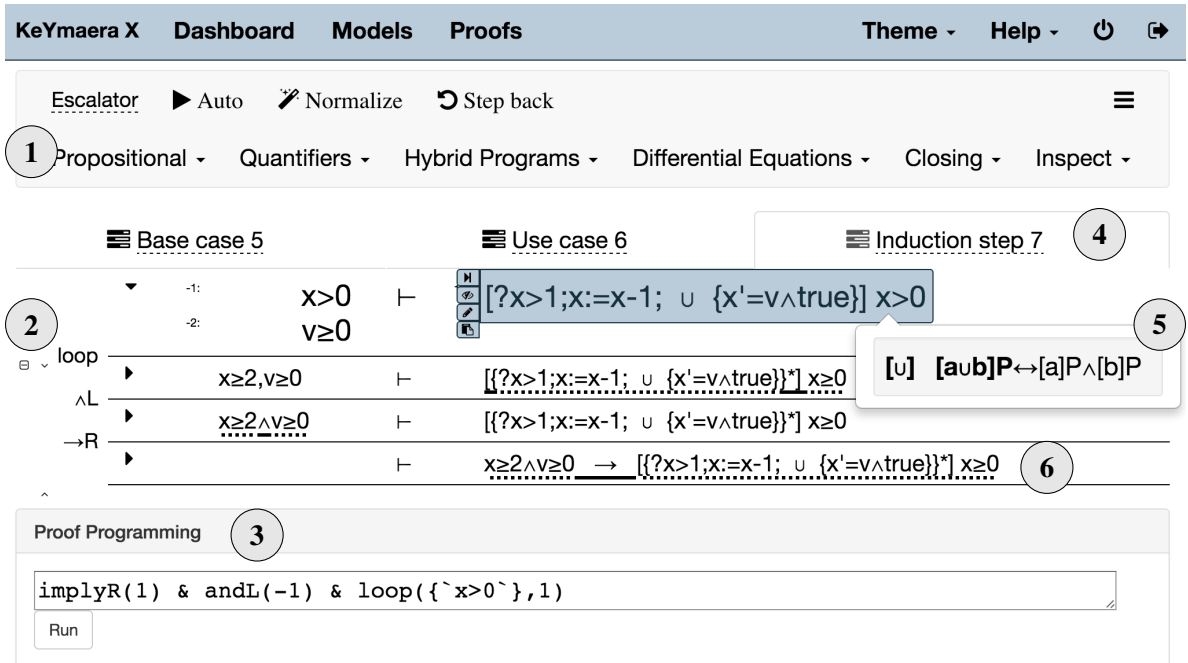

\rref{fig:overview} shows a screenshot of the \KeYmaeraX user interface with user interface elements annotated by their interaction purpose.
We discuss these design choices in more detail in the following paragraphs.

\subsection{Familiarity}

\begin{figure}[t]
\centering
	\subfloat[Sequent proof with step \(\limply\textsf{R}\); the position where \(\limply\textsf{R}\) was applied in the original formula (below the horizontal sequent line) is highlighted with a dotted line, the specific operator with a solid line. \label{fig:sequentproof}]{\includegraphics[width=\columnwidth]{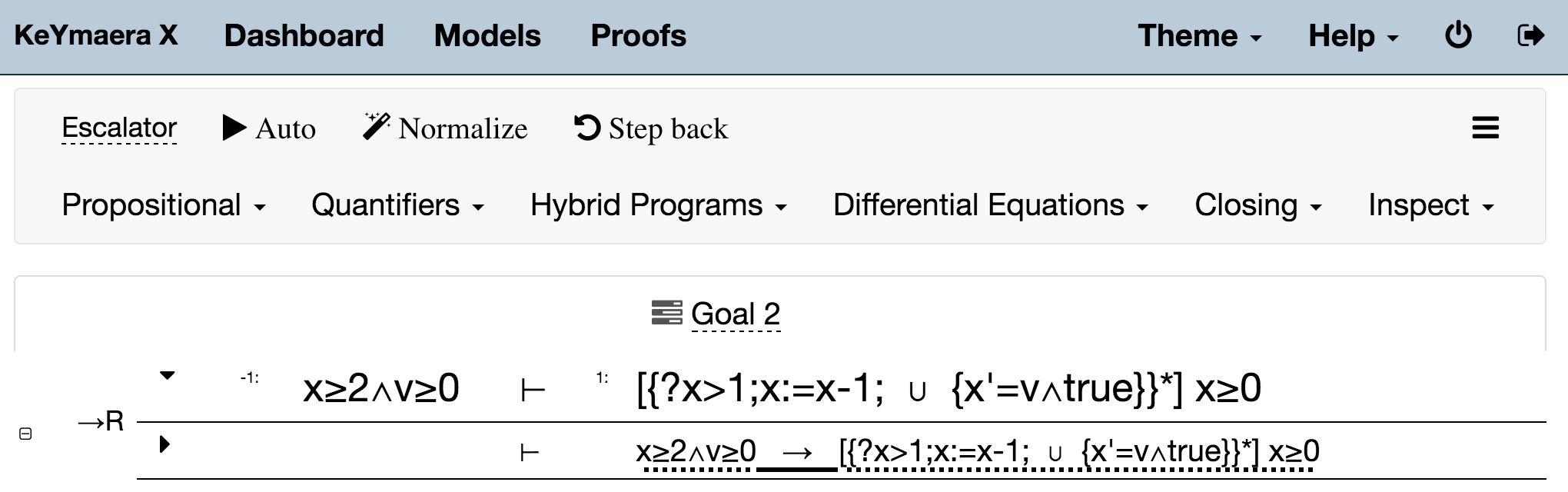}}
    \\
    \subfloat[Axiom \(\dbox{\cup}\)\label{fig:choiceb}]{\includegraphics[width=.2\columnwidth]{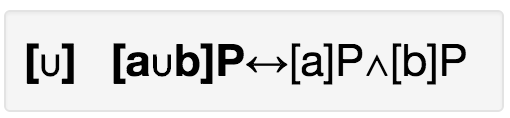}}~
	\subfloat[Sequent rule \(\land\textsf{L}\)\label{fig:andL}]{\includegraphics[width=.35\columnwidth]{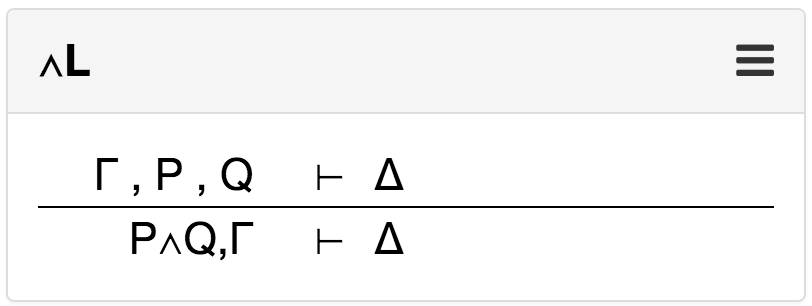}}~
	\subfloat[Sequent rule \textsf{loop}\label{fig:loop}]{\includegraphics[width=.42\columnwidth]{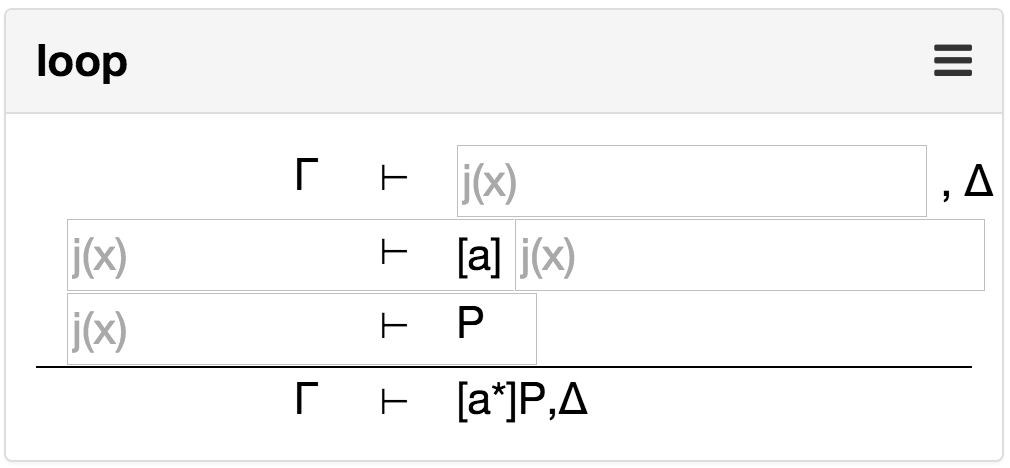}}
\caption{Sequent proof, axioms, and proof rules rendered with standard notation.}
\label{fig:sequent}
\end{figure}

The \KeYmaeraX prover kernel implements Hilbert-style proofs by uniform substitution from a small set of locally sound axioms \cite{DBLP:journals/jar/Platzer16} together with a first-order sequent calculus \cite{DBLP:journals/jar/Platzer08}.
The user interface of \KeYmaeraX presents proofs in sequent form as in \rref{fig:sequent}, which enables users to equivalently read the logical transformations as either sequent proof rule uses \cite{DBLP:journals/jar/Platzer08} or axiom uses \cite{DBLP:journals/jar/Platzer16}.
The rendering is consistent with the notation used in \rref{sec:preliminaries} and in the \emph{Foundations of Cyber-Physical Systems} course \cite{DBLP:conf/cpsed/Platzer13}, which should make it easy to switch between proof development on paper and proving in \KeYmaeraX.
Proof suggestions for tactics are rendered by their primary nature either as axioms (e.\,g., the \dL axiom \(\dbox{\cup}\), see \rref{fig:choiceb}), proof rules (e.\,g., the propositional proof rule \(\land\textsf{L}\), see \rref{fig:andL}), or proof rules with input (e.\,g., the sequent proof rule \textsf{loop}, which requires input, combines multiple axioms in a tactic to implement a proof rule, and creates multiple subgoals, see \rref{fig:loop}).

The hypothesis is that the ability to work from a small number of reasoning principles, which is crucial for a small prover core, helps human understanding as well.
The experience with the \emph{Foundations of Cyber-Physical Systems} course \cite{DBLP:conf/cpsed/Platzer13}, in which \KeYmaera and \KeYmaeraX have been used, suggests that equivalence axioms indeed make it easier for students to understand reasoning principles than explicit sequent proof rules \cite{DBLP:journals/jar/Platzer08}, which obscure inherent dualities for novices.
For example, the equivalence \(\boldsymbol{\dbox{\pchoice{a}{b}}{P}} \lbisubjunct \dbox{a}{P} \land \dbox{b}{P}\) characterizes nondeterministic choices under all circumstances.
Its conjunction $\land$ and the rules for \(\land\) make it apparent that such nondeterministic choices will branch in the succedent but not in the antecedent:
\[
\linfer[choiceb+andR]
{\lsequent{\Gamma}{\dbox{a}{P},\Delta}
&\lsequent{\Gamma}{\dbox{b}{P},\Delta}}
{\lsequent{\Gamma}{\dbox{\pchoice{a}{b}}{P},\Delta}}
\qquad\qquad
\linfer[choiceb+andL]
{\lsequent{\Gamma, \dbox{a}{P}, \dbox{b}{P}} {\Delta}}
{\lsequent{\Gamma, \dbox{\pchoice{a}{b}}{P}} {\Delta}}
\]
Of course, it is exactly the same reasoning principle either way, but understanding the direct sequent proof rules still requires two logical principles at once compared to the single axiom.

\subsection{Traceability}

When switching from automated mode to user guidance, it is important to visualize just enough contextual information about the open proof goals to understand where the present subgoals came from and how they relate to the proof of the ultimate conclusion.
\KeYmaeraX shows local views of proofs that illustrate the way how the current question came about and how it is related to the proof of the conclusion.
The idea is that this enables traceability and gives local justifications while limiting information to the presently relevant part of the proof.

\paragraph{Visualizing the proof state.}
Visualizing the entire proof tree that unfolds from the original conjecture as tree root is not a viable option, since proofs in \dL unfold into many branches and therefore easily exceed the screen width when rendered in a single tree.
Even simple models, such as the escalator example with only four branches (induction base case, induction use case, and stepping down plus moving upwards in the induction step) become hard to navigate and keep track of when both horizontal and vertical scrolling is needed.
\KeYmaeraX, therefore, renders a proof tree in sequent deduction paths from the tree root representing the original conjecture at the bottom of the screen to a leave representing an open goal at the top, see \rref{fig:overview}.


The sequent deduction paths are arranged in tabs; branching occurs when a deduction step has more than one premise, so that premises are spread over multiple tabs and interconnected with links between the tabs.

\paragraph{Proof navigation.} Users can focus solely on the open goal by collapsing the entire deduction path (\(\boxminus\)), or they can keep some part of the proof structure visible, e.\,g., by collapsing only between branching points in the proof, see \rref{fig:overview}. 
Triangles left of the sequent path identify groups: when uncollapsed, down/up arrows (\(\asymp\)) indicate the group borders; when collapsed (\(\rangle\)), the steps in a group are abbreviated into ``\ldots''.
Additionally, sequents can be expanded over multiple lines (\(\blacktriangledown\), one formula per line) or collapsed into a single line (\(\blacktriangleright\)).




When proof automation hands over to the user, the topmost line in a sequent deduction path represents an open goal.
This means that either the goal cannot be proved at all because the model is wrong, or it is not yet proved because user guidance is needed.
In the former case, the counterexample tool allows users to find concrete values that make all formulas on the left of the sequent true but violate all formulas on the right.
In the latter case, users are interested in what to do next (see \rref{sec:tutoring}).

%
%

\subsection{Tutoring}
\label{sec:tutoring}

\paragraph{Suggesting possible proof steps.}

\begin{figure}[b!]
\begin{tikzpicture}
\node at (0,0) {
\includegraphics[width=.97\columnwidth]{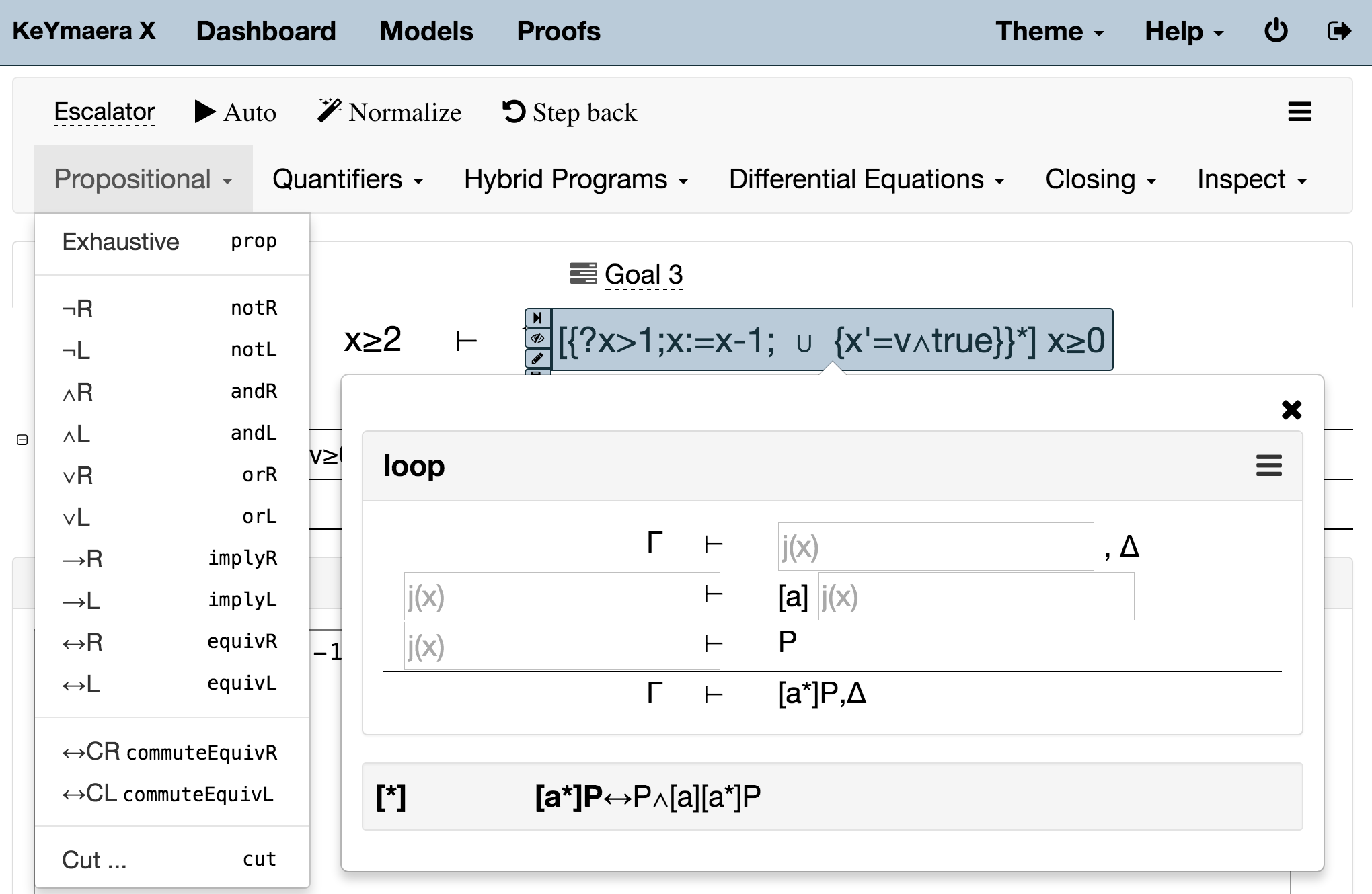}
};
\node[draw,fill=black!10,circle] at (-7.6,3.5) {\bf{1}};
\node[draw,fill=black!10,circle] at (6.6,1) {\bf{2}};
\node[draw,fill=black!10,circle] at (6.6,-1.7) {\bf{3}};
\node[draw,fill=black!10,circle] at (6.6,-4.1) {\bf{4}};
\end{tikzpicture}
\caption{
Two ways of asking \KeYmaeraX for help:
\textcircled{1} Know what to do, but not sure where \(\leadsto\) search tactics in the toolbox menu instruct \KeYmaeraX to apply a specific tactic to the first applicable formula.
\textcircled{2} Know where, but not sure what to do \(\leadsto\) tactic suggestion in the context menu of a formula: \textcircled{3} tactic with input; \textcircled{4} apply an axiom.
}
\label{fig:tacticpopover}
\end{figure}
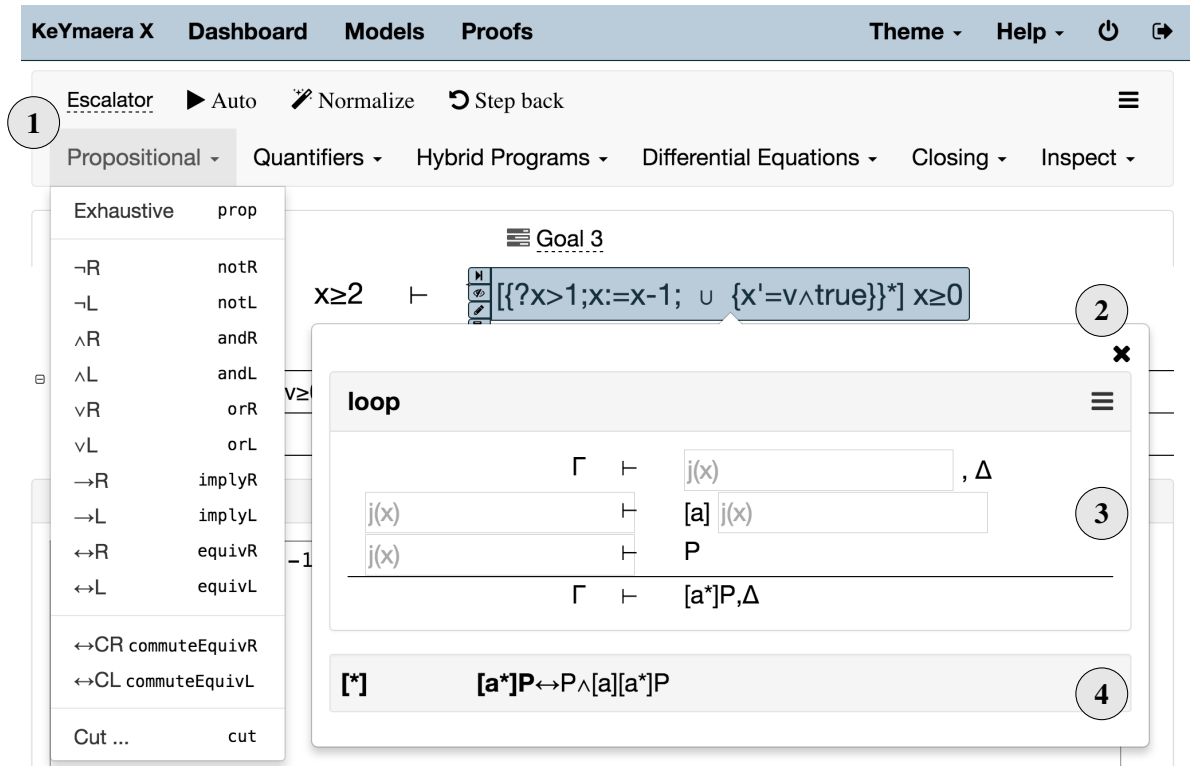

\KeYmaeraX analyzes the shape of formulas to suggest proof steps on demand.
A tactic comes with a description of the shape of its conclusion (must match the current open goal so that the tactic is applicable), a description of the premises that remain to be proved after applying the tactic, and a description of required user input (such as invariants for loops).
Such meta-information allows tactic suggestion in two different flavors, as depicted in \rref{fig:tacticpopover}:
\begin{itemize}
\item When users know where to continue with the proof (i.\,e., exactly on which formula or term or part thereof), \KeYmaeraX displays a dialog with important applicable tactics and their required input, which resembles the user interaction of KeY \cite{DBLP:journals/sosym/AhrendtBBBGHMMRSS05,DBLP:conf/vstte/AhrendtBBBGGHHHKMSSU14} and its hybrid systems descendant \KeYmaera~\cite{DBLP:conf/cade/PlatzerQ08}.
\item When users know what to do next, \KeYmaeraX searches for formulas or terms where the desired tactic is applicable.
\end{itemize}

For example, when pointing to a loop, \KeYmaeraX suggests the induction tactic \textsf{loop} as well as loop unrolling $\dibox{\prepeat{}}$ as shown in \rref{fig:tacticpopover}.
The tactic \textsf{loop} requires a loop invariant \(j(x)\) as input and produces three branches that remain to be proved to conclude safety of the program with the loop. 
Loop unrolling \(\dibox{\prepeat{}}\) turns a formula of the shape in boldface (left-hand side $\boldsymbol{\dbox{\prepeat{\alpha}}{P}}$ of the equivalence) into a formula of the right-hand side $P\land\dbox{\alpha}{\dbox{\prepeat{\alpha}}{P}}$ of the equivalence.

For reasons of flexibility, \KeYmaeraX supports contextual reasoning to apply axioms deeply nested inside a formula, not only on the top-level operator as in \rref{fig:tacticpopover}.
As a consequence, users may have many options to work on a single formula.
In the induction step of the escalator example \[\lsequent{{}}{x{>}0 \limply \dbox{\pchoice{(\ptest{x{>}1};\humod{x}{x-1})}{\D{x}=v}}x{>}0}\] we used \(\limply\textsf{R}\) top-level to get \(\lsequent{x{>}0}{\dbox{\pchoice{(\ptest{x{>}1};\humod{x}{x-1})}{\D{x}=v}}x{>}0}\) as a next goal. 
Alternatively, the non-deterministic choice tactic \(\dbox{\cup}\) on the right-hand side of the implication would result in a conjunction \(\lsequent{{}}{x{>}0 \limply \dbox{\ptest{x{>}1};\humod{x}{x-1}}x{>}0 \land \dbox{\D{x}=v}x{>}0}\).

For novice users, it is often easiest to focus on the top-level operator and work on formulas outside-in, i.\,e., apply \(\limply\textsf{R}\) first and then \(\dbox{\cup}\) next.
But it quickly becomes more convenient to apply proof rules in any order anywhere in the middle of the formulas to follow whatever line of thought the user may have in mind.
Such proofs in arbitrary order also often reduce the branching or repetition of proof steps in different branches.

\paragraph{Tactic extraction.}

\begin{figure}[t!]
\begin{tikzpicture}
\node at (0,0) {
\includegraphics[width=.97\columnwidth]{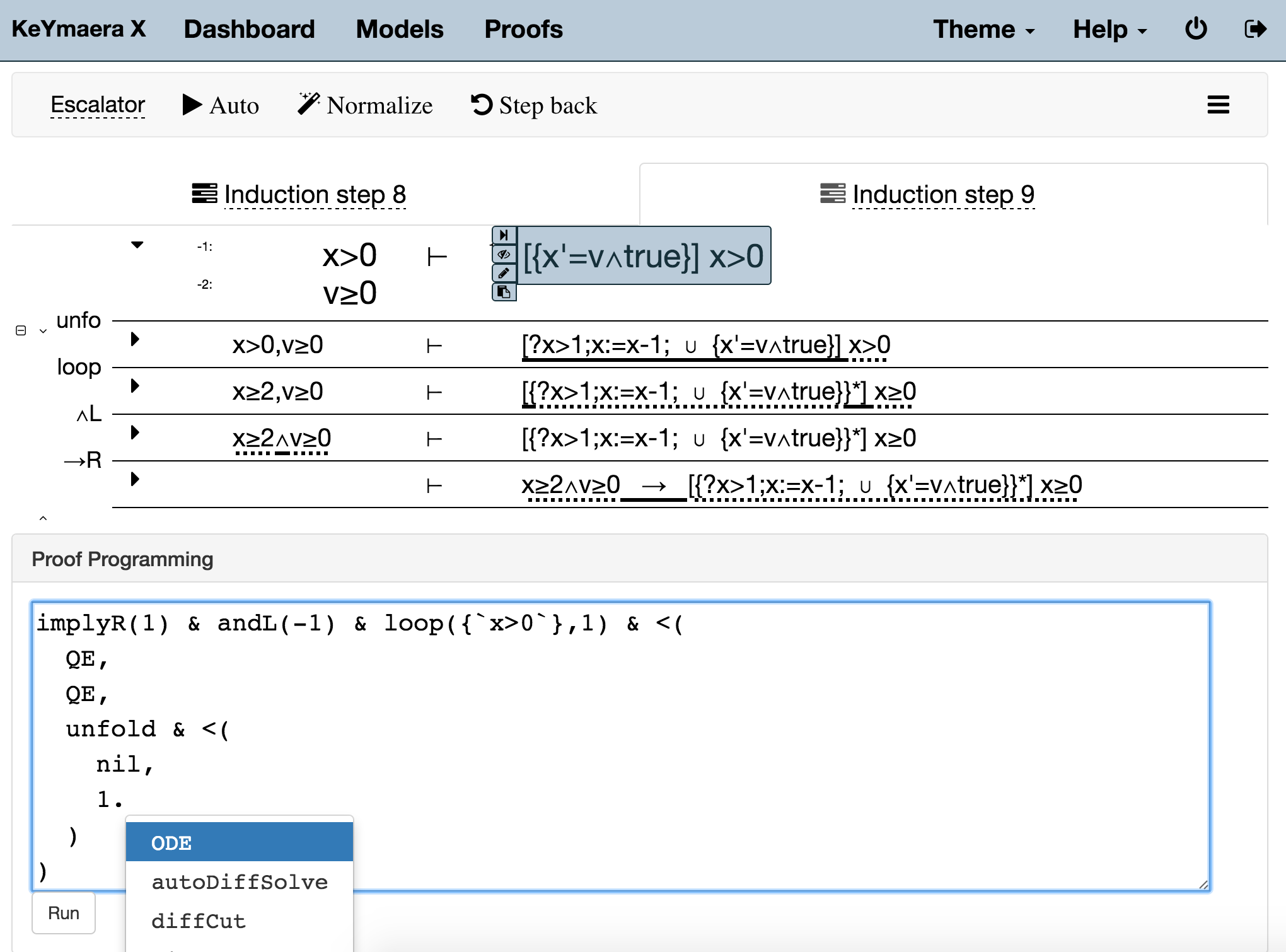}
};
\node[draw,very thick,rectangle,minimum width=0.7cm,minimum height=0.3cm] (seqimply) at (-6.8,0.2) {};
\node[draw,very thick,rectangle,minimum width=1.8cm,minimum height=0.4cm] (tacimply) at (-6.5,-1.8) {};
\draw[dashed,very thick] (seqimply) -- (tacimply);
\node[draw,very thick,rectangle,minimum width=0.7cm,minimum height=0.3cm] (seqand) at (-6.8,0.78) {};
\node[draw,very thick,rectangle,minimum width=1.6cm,minimum height=0.4cm] (tacand) at (-4.4,-1.8) {};
\draw[dashed,very thick] (seqand) to [bend left] node[anchor=west,pos=0.85] {tactic extraction} (tacand);
\node (seqpos) at (1.5,2.7) {};
\node[draw,very thick,rectangle,minimum width=0.5cm,minimum height=0.4cm] (tacpos) at (-6.5,-3.9) {};
\node[right=12.5cm of tacpos] (intermediatepos) {};
\draw[dashed,very thick] (seqpos) -| (intermediatepos)  -- node[fill=white] {position highlighting} (tacpos);
\end{tikzpicture}
\caption{Tactic editor below sequent view. The tactic is extracted automatically from the point-and-click interaction: the proof closed both the induction base case and the induction use case by quantifier elimination \textsf{QE}; the sequent view shows two open goals in the induction step, which result from using \textsf{unfold} (splits the choice, handles the test and the assignment, but stops at the differential equation).
The tactic editor displays tactic suggestions akin to the tactic popover in the sequent view.
Here, it display suggestions for the formula at position ``1'', which is also highlighted in the sequent view.}
\label{fig:tacticeditor}
\vspace{-3ex}
\end{figure}

In order to conduct proofs effectively, interactive theorem provers typically ship with extensive tactic libraries (e.\,g., Coq \cite{Coq:manual}, Isabelle \cite{DBLP:books/sp/NipkowPW02}), accompanied with library documentation and examples to facilitate learning.
Still, extensive tactic libraries incur a steep learning curve. 
We conjecture that observing how tactics evolve while doing a proof (similar to observing an expert) can reduce the steep learning curve associated with extensive tactic libraries.
\KeYmaeraX automatically generates tactics that correspond to the point-and-click interaction that the user performed and displays these tactics below the graphical sequent view, see \rref{fig:tacticeditor}.

The graphical sequent view and the tactic editor operate on the same proof, so that users can switch between both interaction concepts as they see fit.
As a rule of thumb, the sequent view is designed for conducting specific steps at specific positions (e.\,g., use a specific equality \(x=y\) to rewrite \(x\) into \(y\) in some other term), while the tactic editor is intended for describing proof search (e.\,g., repeat some step exhaustively \(\land\textsf{L('L)*}\) or follow a high-level proof strategy such as \textsf{unfold \& ODE \& QE}).

\subsection{Flexibility}

To allow users to reason in flexible ways and reduce manual proof effort, \KeYmaeraX supports proof steps by pointing at formula parts and lets users transform and abbreviate formulas as well as execute tactic scripts.
As underlying technique for proof-by-pointing and flexible reasoning anywhere in formulas, \KeYmaeraX provides contextual reasoning \textsf{CQ} and \textsf{CE} \cite{DBLP:journals/jar/Platzer16}, so that questions about contextual equality or equivalence in a context \(C\{\dots\}\) reduce to reasoning about arguments as follows.

\irlabel{CE|CE}
\irlabel{CQ|CQ}
\irlabel{CMon|CMon}

\begin{minipage}[b]{0.33\linewidth}\centering
\begin{sequentdeduction}[array]
\linfer[CQ]
{\lsequent{{}}{f() = g()}}
{\lsequent{{}}{C\{p(f())\} \lbisubjunct C\{p(g())\}}}
\end{sequentdeduction}
\end{minipage}
\hspace{1em}%
\begin{minipage}[b]{0.33\linewidth}\centering
\begin{sequentdeduction}[array]
\linfer[CE]
{\lsequent{{}}{P \lbisubjunct Q}}
{\lsequent{{}}{C\{P\} \lbisubjunct C\{Q\}}}
\end{sequentdeduction}
\end{minipage}

\noindent When contextual reasoning is combined with uniform substitution \textsf{US} and axiom lookup, axioms can be applied inside formulas, which enables proof by pointing at the desired formula part.
Often, the next proof step follows unambiguously from just the shape of the pointed formula part by indexing \cite{DBLP:journals/jar/Platzer16}, so that the proof advances without further user input.

\irlabel{us|US}
\irlabel{axiomAssignb|ax $[:=]$}
\irlabel{cutAt|cut}
\begin{sequentdeduction}[array]
\linfer[cutAt]
{
\lsequent{{}}{y>2 \land 7>5}
!
\linfer[axiomAssignb]
{\lclose}
{
\linfer[us]
{\lsequent{{}}{\phantom{y>2 \land\,}\dbox{\humod{x}{f()}}p(x) \lbisubjunct p(f())}}
{
\linfer[CE]
{\lsequent{{}}{\phantom{y>2 \land {}}\dbox{\humod{x}{7}}x>5 \lbisubjunct \phantom{y>2 \land {}}7>5}}
{\lsequent{{}}{y>2 \land \dbox{\humod{x}{7}}x>5 \lbisubjunct y>2 \land 7>5}
}
}
}
}
{\lsequent{{}}{y>2 \land \underbrace{\dbox{\humod{x}{7}}x>5}_{\mathclap{\text{unifies with underlined key in axiom } \underline{\dbox{\humod{x}{f()}}p(x)} \lbisubjunct p(f()) ~\leadsto~ \text{replace with }p(f())\text{, which is }7>5}}}
}
\end{sequentdeduction}

Note that side branches with contextual reasoning, uniform substitution, and axiom lookup (as in the proof above) close fully automatically.
Hence, the user interface displays only the result of applying axioms by pointing, while it hides the minutia of the side deductions from the user as in the following example.

\irlabel{useAt|useAt}
\begin{sequentdeduction}[array]
\linfer[useAt]
{\lsequent{{}}{P \limply (\dbox{\D{x}=2}x \geq 5) \lor Q}}
{\lsequent{{}}{P \limply (\underbrace{\dbox{\pchoice{\D{x}=2}{\humod{x}{5}}}x \geq 5}_{\dbox{\beta}B \limply (\underline{\dbox{\pchoice{\alpha}{\beta}}B} \lbisubjunct \dbox{\alpha}B)}) \lor Q}}
\end{sequentdeduction}

\subsection{Experimentation}

The prover kernel and the prover interface are strictly separated, to the extent that the prover kernel only knows about making single deduction steps and combining them, but is completely oblivious of organizing these steps in a tree structure.
For soundness, it suffices to know that any step between the original conjecture and the current subgoals must have been done in the prover kernel. 
Intermediate steps are only necessary to repeat a proof from the original conjecture, and can therefore be tracked outside the prover kernel.
As a result, proof organization features (such as undoing proof steps, pruning the proof tree) can be implemented conveniently in the user interface without affecting soundness.

The separation between the prover core and the tactics becomes especially useful when adding compiled tactics to the server-side implementation.
Complementing the interpreted tactics from the web-based user interface, server-side tactics can base on a rich implementation language (Scala) to try proof steps that are only sound in the usual cases and just rely on the prover core to catch when they happen to be applied in one of the unsound corner cases.

\section{Evaluation Concept}

Compared to its predecessor \KeYmaera \cite{DBLP:conf/cade/PlatzerQ08}, the clean-slate implementation \KeYmaeraX introduces significant changes in user interaction. 
Although the user interaction changes are based on informal feedback from \KeYmaera users and our own observations on how students used \KeYmaera, it still remains to be checked by an experimental user study which style of user interaction is more effective.
Additionally, \KeYmaeraX introduces new interaction concepts (e.\,g., tactic programming and tactic recording), which seem promising for experienced users to conduct large proofs, but are not yet backed by evidence that indeed prover interaction is improved compared to only mouse-based interaction.
Similar user-related research questions were addressed in a recent controlled user experiment \cite{DBLP:conf/kbse/0002HB16}, which followed methods from empirical software engineering \cite{DBLP:books/daglib/0029933} to compare two very different user interfaces of the KeY theorem prover \cite{DBLP:journals/sosym/AhrendtBBBGHMMRSS05,DBLP:conf/vstte/AhrendtBBBGGHHHKMSSU14}.
The obtained results are encouraging pointers towards controlled user experiments being an appropriate method for testing theorem prover interaction.

Such controlled experiments, however, require a large number of participants with varying experience and deliberately exposing them to different user interaction concepts.
In order to avoid adverse effects from pre-assigned  interaction styles, we propose gathering data from normal operation on how often users rely on certain interaction patterns (e.\,g., clicking vs. automated tactic), how often they use a certain functionality during a proof, and how successful they were, even if that might lead to selection bias.
Learnability of theorem provers can be measured based on cognitive dimensions \cite{DBLP:conf/ct/BlackwellBCGGKKLNPRRWY01}, such as visibility (how easy are relevant steps accessed), juxtaposability (different notations side-by-side), viscosity (resistance to change), premature commitment (to an order how to do steps), error-proneness (how likely are errors), and consistency (similar information presented in similar ways).
Theorem provers, especially in education and also in industrial applications, should have high visibility and juxtaposability, high consistency, and low viscosity \cite{Kadoda1999}.
The following metrics should be easily recordable from \KeYmaeraX without changing user interaction and give insights into cognitive dimensions.

\begin{description}
\item[Interaction concept] Number of proof steps by clicking/tactic programming/automated proof search.
Supporting evidence: number of proof steps at top-level/inside formulas; for clicking: number of steps executed by pointing/from tactic suggestion. Related to visibility and premature commitment.
\item[Functionality] Number of undo operations, length of pruned deduction paths with distance to next branching point above and below, number of find counterexample operations.
Number of interactions in pruned proofs.
Related to viscosity and error-proneness.
\item[Time] Proof duration (including estimates of user time and number of reproof attempts)
\item[Trends] Compare trends as students gain experience and examples become more complex
\end{description}

\noindent We believe this data should enable a study that tests the following hypotheses.
\begin{description}
\item[Interaction preference] Novice users prefer automated proof search over clicking over tactic programming. 
Novice users prefer working top-level  over working inside formulas.
Experienced users balance interaction.
\item[Functionality] Novice users either undo short paths or entire proofs and end up with more open branches.
Experienced users undo branching and exercise branching control techniques.
\item[Trends] Automated usage drops with increased example complexity and student experience, while use of tactic programming increases; the focus on applying steps top-level drops while applying steps inside formulas increases with experience.
\item[Influence] Students that engage more direct control over proofs also for simpler examples are more productive in conducting proofs of complex cases than students who relied on full automation earlier on.
\end{description}

The quantitative insights from these metrics could be augmented with qualitative evaluation techniques, as demonstrated being effective for theorem provers, e.\,g., with focus groups \cite{DBLP:conf/sefm/BeckertGB14} (users subjectively express their perception of or opinion on the user interface), with questionnaires \cite{DBLP:conf/cade/BeckertG12}, or by co-operative evaluation \cite{Jackson1997} (users verbalize their interactions while using the theorem prover).



\section{Related Work}

The verification landscape spans a wide variety of approaches from automated theorem provers and reachability analysis tools to interactive theorem provers.

Automated theorem provers (e.\,g., Vampire \cite{DBLP:conf/cav/KovacsV13}) and reachability analysis tools (e.\,g., SpaceEx \cite{DBLP:conf/cav/FrehseGDCRLRGDM11}) strive for fully automatic verification without user interaction, but their scope is inherently limited in cases that are not semidecidable, such as hybrid systems.

Auto-active verifiers (e.\,g., Dafny \cite{DBLP:conf/icse/Leino04,DBLP:journals/corr/LeinoW14}, AutoProof \cite{DBLP:conf/tacas/TschannenFNP15}) put the model or code first and hide the verification engine, but support user guidance through annotations in the code.
The basic idea is to make verification an integral part of (software) development that should be performed in the background by an IDE, much like background compilation.
The downside of such an approach is that the verification steps are hidden entirely from the user, which can make it hard to resolve proofs with additional annotations when the verifier is stuck because the proof state and the verifier's working principles are opaque.
The KeY interactive verification debugger \cite{DBLP:conf/kbse/0002HB16a} combines code annotations with interactive verification in KeY \cite{DBLP:journals/sosym/AhrendtBBBGHMMRSS05,DBLP:conf/vstte/AhrendtBBBGGHHHKMSSU14} to supplement manual proofs when automated proving fails at some goals.

Interactive theorem provers, such as Coq \cite{Coq:manual} and Isabelle \cite{DBLP:books/sp/NipkowPW02}, primarily interact with users through tactic scripts, such as structured proofs in Isabelle/Isar \cite{DBLP:conf/types/Nipkow02}. 
Their user interfaces (e.\,g., CoqIDE \cite{Coq:manual}, ProofGeneral \cite{DBLP:conf/mkm/AspinallLW07}, jEdit \cite{DBLP:conf/aisc/Wenzel12}) focus on text editing support for writing tactics and let users inspect the proof state and open goals by placing the cursor in the tactic script.
Navigation with cursors introduced a limited form of proof by pointing \cite{DBLP:journals/entcs/AspinallL04a} to fold or unfold equations.
\KeYmaeraX advances proof-by-pointing to transform all or parts of a formula following the shape of an axiom or fact.
PeaCoq\footnote{\url{http://goto.ucsd.edu/peacoq/}} aims to make the proof state more accessible to users by providing a proof tree.

\KeYmaeraX combines concepts from automated, auto-active, and interactive theorem proving: it comes with fully automated proof search tactics for hybrid systems of limited scope (i.\,e., when a loop invariant can be found, a symbolic solution of a differential equation can be found to serve as an oracle for differential invariants, and the resulting arithmetic is tractable); it supports annotations for loop invariants and differential invariants, since both serve as model documentation as well as proof guidance; and finally, it allows users to conduct and finish proofs themselves with tactics and a graphical user interface.


\section{Conclusion}

The user interface of \KeYmaeraX is based on the principles of familiarity, traceability, tutoring, flexibility, and experimentation.
It supports several different user interaction styles to make progress in proofs.
The web-based user interface applies proof steps when clicking on formulas in the sequent view, and it batch-runs proof steps by automated search tactics as well as through tactic programming.

Future work includes evaluation in controlled user experiments, as described in the evaluation concept.
On the basis of user studies, we expect to gain insights into how to best teach hybrid systems theorem proving, tactic programming, tactic generalization, and proof search.
We are working on improved proof exploration, e.g., with timeouts (e.\,g., allow \textsf{QE} 5s to close; if it does not close within the budgeted time, try something else).
Failed proof attempts or expired timeouts need a robust approach to making proofs portable and repeatable.


\paragraph{Acknowledgments}
The authors thank Nathan Fulton, Brandon Bohrer, Jan-David Quesel, Ran Ji, and Marcus V\"olp for their support in implementing \KeYmaeraX, Sarah Loos, Jean-Baptiste Jeannin, Jo{\~a}o Martins, Khalil Ghorbal, and Sarah Grebing for feedback and discussions on the user interface, as well as the anonymous reviewers for feedback on the manuscript.

\noindent This material is based on research sponsored by DARPA under agreement DARPA FA8750-12-2-0291.

\bibliographystyle{eptcs}
\bibliography{webui}


\clearpage
\appendix

\end{document}